\documentclass[12pt]{article}

\begin{document}

\begin{center}
{\bf Modified wave equation for spinless particles and its solutions in an external magnetic field} \\
\vspace{5mm} S. I. Kruglov\\
\vspace{3mm}
\textit{Department of Chemical and Physical Sciences,\\ University of Toronto at Mississauga,\\
3359 Mississauga Rd. North, Mississauga, Ontario, Canada L5L 1C6} \\
\vspace{5mm}
\end{center}

\begin{abstract}
The wave equation for spinless particles with the Lorentz violating term is
considered. We formulate the third-order in derivatives wave equation leading to the modified dispersion relation. The first-order formalism is considered and the density matrix is obtained. The Schr\"{o}dinger form of equations is presented and the quantum-mechanical Hamiltonian is found. Exact solutions of the wave equation are obtained for particles in the constant and uniform external magnetic field.
The change of the synchrotron radiation radius due to quantum gravity corrections is calculated.
\end{abstract}

\section{Introduction}

The deformed dispersion relations, which result in the Lorentz violation, can be caused by quantum gravity \cite{Amelino}, \cite{Coleman}, \cite{Gambini}, \cite{Amelino1}, \cite{Amelino2},
\cite{Kowalski}, \cite{Smolin}, \cite{Smolin1}, \cite{Amelino3} \cite{Lee}, \cite{Myers}. According to \cite{Amelino1}, \cite{Amelino2}, we consider the special case of the modified dispersion relation (the speed of light in vacuum $c$ equals unit in our notations):
\begin{equation}
p_0^2=\textbf{p}^2+m^2-Lp_0\textbf{p}^2,
\label{1}
\end{equation}
where $p_0$ is an energy and $\textbf{p}$ is a momentum of a particle and $L$ is a parameter with the dimension of ``length". The positive value of $L$ ($L>0$) corresponds to the subluminal propagation of  particles. We imply that the last term in Eq.(1), violating the Lorentz symmetry, is due to quantum gravity corrections, and $L$ is of the order of the Planck length $L_P=M_P^{-1}$ ($M_P=1.22\times 10^{19}$ GeV is the Planck mass). The modified dispersion relation (1) appears in space-time foam Liouville-string models \cite{Ellis}, \cite{Ellis1}.
Constrains on quantum gravity corrections were estimated from the Crab Nebula synchrotron
radiation \cite{Jacobson}, \cite{Jacobson1}, \cite{Ellis2}. The modified dispersion relation (1) can be motivated within Doubly Special Relativity (DSR) \cite{Amelino1}, \cite{Amelino2},
\cite{Kowalski}, \cite{Smolin}, \cite{Smolin1}, \cite{Amelino3} \cite{Lee}.
The goal of this paper is to describe spinless fields realizing the deformed dispersion relation (1) in third-order and first-order formalisms. Also we obtain equation solutions of free particles and particles in the external magnetic field.

The paper is organized as follows. In Sec.2, we formulate the wave equation
with modified dispersion relation in the third-order and first-order formalisms and obtain the density matrix. The Schr\"{o}dinger form of equations is presented and the quantum-mechanical Hamiltonian is found in Sec.3. In Sec.4 we obtain exact solutions of the wave equation for particles in the constant and uniform external magnetic fields.
The synchrotron radius with quantum gravity corrections is estimated.
A conclusion is made in Sec.5. In Appendix, Sec.6, useful products of the equation matrices are obtained.

The Euclidean metric is explored and the system of units $\hbar =c=1$ is used. Greek letters run 1,2,3,4 and Latin letters run 1,2,3.

\section{Field equation for spinless particle}

\subsection{Wave equation in the first-order formalism}

Let us consider the wave equation for spinless particles:
\begin{equation}
\left(\partial_{\mu}^2-m^2-iL\partial_i^2\partial_t\right)\Phi(x)=0,
\label{2}
\end{equation}
where $\partial_\mu=\partial/\partial x_\mu=\left(\partial/\partial x_i,\partial/(i\partial t)\right)$, $x_0=t$ is a time.
One can treat Eq.(2) as an effective wave equation which takes into account quantum gravity corrections.
The plane-wave solution for positive energy $\Phi(x)=\Phi_0 \exp[i(\textbf{p}\textbf{x}-p_0x_0)]$ to Eq.(2) leads to the modified dispersion relation (1). Eq.(2) is invariant under the rotation group but the last term in (2) violates the invariance under the boost transformations. Thus, the Lorentz symmetry is broken. One can try to interpret Eq.(2) to be invariant under DSR transformations but we do not prove this because DSR is formulated in the momentum space and there is not the consistent formulation of the model in the position space yet. Therefore, it is possible to consider Eq.(2) as an equation with the Lorentz violating term and introducing preferred frame effects. We will call the $L$ a deformation parameter.

To present the higher derivative equation (2) in the first-order formalism, we follow the method of \cite{Kruglov}. Let us introduce the system of first order equations which are equivalent to Eq.(2)
\[
\partial _\mu \Psi _\mu +m \Phi+\partial_4\widetilde{\Phi}=0,
\]
\begin{equation}
\partial_\mu\Phi+m\Psi _\mu=0,
\label{3}
\end{equation}
\[
L\partial _m \Psi _m -\widetilde{\Phi}=0.
\]
Indeed, replacing $\Psi_\mu$ and $\widetilde{\Phi}$ from Eqs.(3) into the first equation of (3), one obtains Eq.(2).  The fields $\Psi_\mu$, $\Phi$, $\widetilde{\Phi}$ possess the same dimension.
It is convenient to introduce the wave function
\begin{equation}
\Psi (x)=\left\{ \Psi _A(x)\right\} =\left(
\begin{array}{c}
\Phi(x)\\
\Psi_\mu (x)\\
\widetilde{\Phi}(x)
\end{array}
\right),
\label{4}
\end{equation}
and index runs $A=(0,\mu ,\widetilde{0})$, $\Psi_0=\Phi$, $\Psi_{\widetilde{0}}=\widetilde{\Phi}$.
With the help of the elements of the entire matrix algebra $\varepsilon
^{A,B}$, with properties \cite{Kruglov1}
\begin{equation}
\left( \varepsilon ^{M,N}\right) _{AB}=\delta _{MA}\delta _{NB},
\hspace{0.5in}\varepsilon ^{M,A}\varepsilon ^{B,N}=\delta
_{AB}\varepsilon ^{M,N},
\label{5}
\end{equation}
where $A,B,M,N=(0,\mu ,\widetilde{0})$, the system of equations (3) may be
presented in the matrix form as follows:
\[
\biggl[\partial _\mu \left(\varepsilon ^{\mu,0}+ \varepsilon
^{0,\mu}+\delta_{\mu 4}\varepsilon ^{0,\widetilde{0}}-mL\delta_{\mu m}\varepsilon ^{\widetilde{0},m}\right)
\]
\vspace{-8mm}
\begin{equation}
\label{6}
\end{equation}
\vspace{-8mm}
\[
+ m\left(\varepsilon ^{0,0}+\varepsilon ^{\mu,\mu}+
\varepsilon ^{\widetilde{0},\widetilde{0}}\right)\biggr] _{AB}\Psi
_B(x)=0 .
\]
with the summation over all repeated indices. We introduce the
$6\times 6$ matrices
\begin{equation}
\beta_\mu=\varepsilon ^{\mu,0}+ \varepsilon
^{0,\mu}+\delta_{\mu 4}\varepsilon ^{0,\widetilde{0}}-mL\delta_{\mu m}\varepsilon ^{\widetilde{0},m},~~~
I_6=\varepsilon ^{0,0}+\varepsilon ^{\mu,\mu}+
\varepsilon ^{\widetilde{0},\widetilde{0}},
\label{7}
\end{equation}
so that
\begin{equation}
\beta_m=\varepsilon ^{m,0}+ \varepsilon
^{0,m}-mL\varepsilon ^{\widetilde{0},m},~~~
\beta_4=\varepsilon ^{4,0}+ \varepsilon
^{0,4}+\varepsilon ^{0,\widetilde{0}}.
\label{8}
\end{equation}
Taking into account these equations Eq.(6) becomes the first-order wave equation
\begin{equation}
\left( \beta _\mu \partial _\mu + m\right)
\Psi(x)=0 , \label{9}
\end{equation}
where we have used the unit $6\times 6$-matrix $I_6$.
We note that the 5-dimensional matrices
\begin{equation}
\beta_\mu^{(0)}=\varepsilon ^{\mu,0}+ \varepsilon
^{0,\mu}
\label{10}
\end{equation}
enter the Lorentz covariant wave equation for scalar particles
\begin{equation}
\left( \beta _\mu^{(0)} \partial _\mu + m\right)
\Psi^{(0)}(x)=0 , \label{11}
\end{equation}
where the wave function reads
\begin{equation}
\Psi^{(0)}(x)=\left(
\begin{array}{c}
\Phi(x)\\
\Psi_\mu (x)
\end{array}
\right).
\label{12}
\end{equation}
Matrices (10) obey the Duffin$-$Kemmer$-$Petiau algebra \cite{Kruglov1}
\begin{equation}
\beta^{(0)}_\mu \beta^{(0)} _\nu \beta^{(0)} _\alpha +\beta^{(0)} _\alpha \beta^{(0)} _\nu
\beta^{(0)} _\mu =\delta _{\mu \nu }\beta^{(0)} _\alpha+\delta _{\alpha \nu
}\beta^{(0)} _\mu . \label{13}
\end{equation}
The Lorentz group generators in the $5$-dimension representation space are given by
\begin{equation}
J_{\mu\nu}=\beta^{(0)}_\mu\beta^{(0)}_\nu-\beta^{(0)}_\nu\beta^{(0)}_\mu=
\varepsilon^{\mu,\nu}- \varepsilon^{\nu,\mu},
 \label{14}
\end{equation}
and obey the commutation relations
\[
\left[J_{\rho\sigma},J_{\mu\nu}\right]=\delta_{\sigma\mu}J_{\rho\nu}+
\delta_{\rho\nu}J_{\sigma\mu}-\delta_{\rho\mu}J_{\sigma\nu}-\delta_{\sigma\nu}J_{\rho\mu},
\]
\vspace{-8mm}
\begin{equation}
\label{15}
\end{equation}
\vspace{-8mm}
\[
\left[\beta^{(0)}_{\lambda},J_{\mu\nu}\right]=\delta_{\lambda\mu}\beta^{(0)}_\nu
-\delta_{\lambda\nu}\beta^{(0)}_\mu.
\]
It should be noted that the field $\widetilde{\Phi}(x)$ in Eqs.(3),(4) is a scalar under the rotation group but is not a scalar under the boost transformations. Therefore, the expression $J_{mn}=\varepsilon^{m,n}-\varepsilon^{n,m}$ can be considered as the rotation group generators in the $6$-dimension representation space of wave functions (4). From Eq.(7), one obtains the commutators
as follows:
\begin{equation}
\left[\beta_4,J_{m n}\right]=0,~~~~\left[\beta_k,J_{m n}\right]=\delta_{k m}\beta_n
-\delta_{k n}\beta_m.
\label{16}
\end{equation}
Eqs.(16) indicate that wave equation (9) is covariant under the group of rotations but the form-invariance of Eq.(9) under the Lorentz transformations is broken due to terms containing the deformation parameter $L$. With the help of products of $\beta$-matrices (see Appendix) I find the algebraic relation
\[
\beta_\mu\left(\beta_\nu\beta_\lambda\beta_\sigma+\beta_\sigma\beta_\lambda\beta_\nu\right)+
\beta_\lambda\left(\beta_\nu\beta_\mu\beta_\sigma+\beta_\sigma\beta_\mu\beta_\nu\right)+
\beta_\nu\left(\beta_\lambda\beta_\sigma\beta_\mu+\beta_\mu\beta_\sigma\beta_\lambda\right)
\]
\[
+\beta_\sigma\left(\beta_\lambda\beta_\nu\beta_\mu+\beta_\mu\beta_\nu\beta_\lambda\right)=
\delta_{\mu\nu}\left(\beta_\lambda\beta_\sigma+\beta_\sigma\beta_\lambda\right)+
\delta_{\lambda\nu}\left(\beta_\mu\beta_\sigma+\beta_\sigma\beta_\mu\right)
\]
\vspace{-8mm}
\begin{equation}
\label{17}
\end{equation}
\vspace{-8mm}
\[
+\delta_{\mu\sigma}\left(\beta_\lambda\beta_\nu+\beta_\nu\beta_\lambda\right)
+\delta_{\sigma\lambda}\left(\beta_\mu\beta_\nu+\beta_\nu\beta_\mu\right)
\]
\[
-mL\biggl[\left(\delta_{\sigma m}\delta_{\nu 4}+\delta_{\nu m}\delta_{\sigma 4}\right)\left(\delta_{m\lambda}\beta_\mu +\delta_{m\mu}\beta_\lambda\right)
+ \left(\delta_{\lambda m}\delta_{\mu 4}+\delta_{\mu m}\delta_{\lambda 4}\right)\left(\delta_{m\sigma}\beta_\nu+\delta_{m\nu}\beta_\sigma\right)\biggr].
\]
It should be noted that, for example, $\delta_{\sigma m}\delta_{m\lambda}\neq \delta_{\sigma\lambda}$ because $\delta_{4 m}\delta_{m4}=0$ but $\delta_{44}=1$. The algebra of $\beta$-matrices (17) is more complicated compared with the Duffin$-$Kemmer$-$Petiau algebra (13).

\subsection{The density matrix}

In the momentum space, for the positive energies $\Psi(x)\sim \exp[i(\textbf{p}, \textbf{x}-p_0x_0)]$ and Eq.(9) becomes
\begin{equation}
\left( i\widehat{p} + m\right) \Psi (p)=0 , \label{18}
\end{equation}
where $\widehat{p}=\beta _\mu p _\mu$. It follows from Eq.(17) that the matrix $\widehat{p}$ obeys the equation as follows (see also Appendix):
\begin{equation}
\widehat{p}^4 -p^2\widehat{p}^2+mLp_4\textbf{p}^2\widehat{p}=0 , \label{19}
\end{equation}
where $p^2=\textbf{p}^2-p_0^2$, $p_4=ip_0$.
With the help of (19), one may prove that the matrix
\begin{equation}
\Lambda = i\widehat{p} + m
\label{20}
\end{equation}
obeys the matrix equation
\begin{equation}
\Lambda^4-4m\Lambda^3+\Lambda^2\left(p^2+6m^2\right)-\Lambda\left(2mp^2+4m^3+imLp_4\textbf{p}^2 \right)
=0. \label{21}
\end{equation}
It follows from Eq.(21) that solutions to Eq.(18) in the form of the projection matrix are given by
\begin{equation}
\Pi=N\left[\Lambda^3-4m\Lambda^2+\Lambda\left(p^2+6m^2\right)-2mp^2-4m^3-imLp_4\textbf{p}^2
\right]. \label{22}
\end{equation}
so that $\left(i\widehat{p} + m\right)\Pi=0$, and $N$ is the normalization
constant. The requirement that $\Pi$ is the projection matrix \cite{Fedorov} gives
\begin{equation}
\Pi^2=\Pi. \label{23}
\end{equation}
From Eq.(23), with the help of Eq.(21), we obtain the normalization constant
\begin{equation}
N=-\frac{1}{m\left(2m^2+Lp_0\textbf{p}^2\right)} .
\label{24}
\end{equation}
Equation (22) for the matrix $\Pi$ can be simplified using Eqs.(19),(24), and the result is
\begin{equation}
\Pi=\frac{i\widehat{p}\left(\widehat{p}^2+im\widehat{p}-Lp_0\textbf{p}^2\right)}
{m\left(2m^2+Lp_0\textbf{p}^2\right)}.
\label{25}
\end{equation}
The density matrix (25) can be used for calculating some processes with scalar particles obeying Eq.(9) in the perturbation theory. Every column of the matrix $\Pi$ is the solution to Eq.(18). The wave function $\Psi(p)$ in Eq.(18) also can be written as $\Psi(p)=\Pi\Psi_0$, where $\Psi_0$ is arbitrary non-zero 6-component vector.

\section{The Schr\"{o}dinger form of the equation}

Introducing interactions of scalar particles under consideration with external electromagnetic fields by replacing $\partial_\mu\rightarrow D_\mu=\partial_\mu-ieA_\mu$ ($A_\mu$ is the vector-potential of the electromagnetic fields, $e$ is the charge of the particle), we rewrite
Eq.(9) as follows:
\begin{equation}
i\beta _4\partial _t\Psi (x)=\biggl (\beta
_mD_m+m+eA_0\beta _4\biggr )\Psi (x).
 \label{26}
\end{equation}
The matrix $\beta_4$ obeys the matrix equation (see Appendix)
\begin{equation}
 \beta _4^4=\beta_4^2 .
\label{27}
\end{equation}
Thus, the matrix $\Sigma=\beta_4^2$ is the projection operator, $\Sigma^2=\Sigma$.
The matrix $\Sigma$, acting on the wave function $\Psi(x)$, retains only the dynamical components $\phi
(x)=\Sigma \Psi(x)$ of the wave function $\Psi(x)$. To separate the dynamical
components of the wave function $\Psi (x)$ from Eq.(26), we consider the projection operator
\begin{equation}
\Omega =I_6-\Sigma = \varepsilon ^{\widetilde{0},\widetilde{0}}+\varepsilon ^{m,m}-\varepsilon ^{4,\widetilde{0}}.
\label{28}
\end{equation}
One can verify that $\Omega^2=\Omega$. Non-dynamical components of the wave function $\Psi (x)$ are defined by $\chi=\Omega\Psi (x)$. After multiplying Eq.(26) by the matrix $\beta_4$,
we obtain the equation
\begin{equation}
i\partial _t \phi(x)=\beta_4\beta
_mD_m\Psi+m\beta_4\Psi+eA_0\phi. \label{29}
\end{equation}
One can use the relation $\Psi(x)=\phi(x)+\chi(x)$ as $\Sigma+\Omega=I_6$. With the help of equations (see Appendix) $\beta_4\Omega=0$, $\beta_4\beta_m\Sigma=0$, we find from Eq.(29)
\begin{equation}
i\partial _t \phi(x)=\beta_4\beta
_mD_m\chi+\left(m\beta_4+eA_0\right)\phi. \label{30}
\end{equation}
Non-dynamical components $\chi(x)$ can be eliminated from Eq.(30). Indeed, multiplying
Eq.(26) by the matrix $\Omega$, and taking into consideration the equality $\Omega \beta_4
=0$, one finds the equation as follows:
\begin{equation}
\Omega \beta_n D_n\Psi +m\chi=0 .
\label{31}
\end{equation}
Eliminating $\chi$ from Eq.(31) and replacing it into Eq.(30), with the help of the relation $\beta_4\beta_m\Omega\beta_n\Omega=0$, we obtain the Schr\"{o}dinger form of the equation
\begin{equation}
i\partial _t\phi (x)=\biggl (
-\frac{1}{m}\beta_4\beta_mD_m\Omega\beta_n D_n+m\beta_4 +eA_0\biggr)\phi(x).
\label{32}
\end{equation}
It is easy to verify that the wave function $\phi$ possesses only two non-zero components:
\begin{equation}
\phi (x)=\left(
\begin{array}{c}
\Phi(x)\\
\Psi_4(x)+\widetilde{\Phi}(x)
\end{array}
\right).
\label{33}
\end{equation}
The wave function (33) corresponds to two states with positive and negative energies
and does not contain auxiliary components. Then Eq.(32) takes the form
\begin{equation}
i\partial _t\phi (x)=\mathcal{H}\phi(x),
\label{34}
\end{equation}
where the Hamiltonian is given by (see Appendix)
\begin{equation}
\mathcal{H}=m\left(\varepsilon^{4,0}+\varepsilon^{0,4}\right)
+eA_0\left(\varepsilon^{0,0}+\varepsilon^{4,4}\right)-\frac{1}{m}\left(\varepsilon^{4,0}
-mL\varepsilon^{0,0}\right)D_m^2.
\label{35}
\end{equation}
Using (5), the quantum-mechanical Hamiltonian becomes
\begin{equation}
\mathcal{H}=\left(
\begin{array}{cc}
eA_0+LD_m^2&m \\
m-(1/m)D_m^2&eA_0
\end{array}\right).
\label{36}
\end{equation}
One can rewrite Eq.(34), with the help of Eqs.(33),(36), in the component form
\[
i\partial _t\Phi (x)= eA_0 \Phi(x)+ m\left(\Psi_4(x)+\widetilde{\Phi}(x)\right)+LD_m^2\Phi(x) ,
\]
\vspace{-7mm}
\begin{equation} \label{37}
\end{equation}
\vspace{-7mm}
\[
i\partial _t\left(\Psi_4(x)+\widetilde{\Phi}(x)\right)=m\Phi(x)+eA_0\left(\Psi_4(x)
+\widetilde{\Phi}(x)\right)-\frac{1}{m}D_m^2\Phi(x) .
\]
One can check that Eqs.(37) can be obtained from Eqs.(3), after the
replacement $\partial_\mu\rightarrow D_\mu$ and the exclusion of non-dynamical
components $\Psi_m (x)=-(1/m)D_m \Phi(x)$. Eqs.(37) and (34) contain only
components with the time derivatives. The Schr\"{o}dinger equation (34) with the Hamiltonian (36)
can be used for solving different quantum mechanical problems. The matrix Hamiltonian (36) for free space ($A_\mu=0$) in the momentum space $\partial_\mu\rightarrow ip_\mu$ obeys the equation
\begin{equation}
\mathcal{H}^2+L\textbf{p}^2\mathcal{H}-\left(\textbf{p}^2+m^2\right)I_2=0,
\label{38}
\end{equation}
where $I_2$ is the unit $2\times 2$-matrix. From Eq.(38), one finds that the eigenvalue of the
Hamiltonian (36) $p_0$ satisfies the dispersion equation (1).

\section{Particle in an external magnetic field}

Introducing the electromagnetic interaction of particles in the standard way by substitution $\partial_\mu\rightarrow D_\mu=\partial_\mu-ieA_\mu$, Eq.(2) becomes \footnote{It should be noted that authors of the paper \cite{Ellis2} chose another coupling of a particle with the potential.}
\begin{equation}
\left[\left(\partial_{\mu}-ieA_\mu\right)^2-m^2-iL\left(\partial_i-ieA_i\right)^2
\left(\partial_t+ieA_0\right)\right]\Phi(x)=0,
\label{39}
\end{equation}
and $t=x_0$ is the time. It is obvious that the gauge invariance of Eq.(39) is preserved under the transformations:
\[
A_\mu(x)\rightarrow A_\mu(x)+\partial_\mu\Lambda(x),~~~\Phi(x)\rightarrow\Phi(x)\exp\left(ie\Lambda(x)\right).
\]
For a uniform and static magnetic field, we can take the 4-potential in the form
\begin{equation}
A_\mu=\left(-\frac{1}{2}x_2H,\frac{1}{2}x_1H,0,0\right),
\label{40}
\end{equation}
and the magnetic field becomes along the $x_3$ axis, $\textbf{H}=(0,0,H)$. For the potential (40) the Lorentz condition $\partial_\mu A_\mu=0$ and Coulomb condition $\partial_m A_m=0$ are satisfied, and $A_0=0$. Let us consider the motion of negative particles, $e=-e_0$, $e_0>0$. Then
Eq.(39) for a spinless particle with the potential (40) reads
\begin{equation}
\left\{\left[\partial_m^2-e_0HL_3-\frac{e_0^2H^2}{4}\left(x_1^2+x_2^2\right)\right]\left(1-iL\partial_t\right)
-\partial_t^2-m^2\right\}\Phi\left(\textbf{x},t\right)=0,
\label{41}
\end{equation}
where the projection operator of the angular momentum on the $x_3$ axis is given by $L_3=i\left(x_2\partial_1-x_1\partial_2\right)$. To solve Eq.(41), we follow very closely to \cite{Kruglov2}, \cite{Kruglov3}. The solution to Eq.(41) can be obtained by the substitution \cite{Sokolov}
\begin{equation}
\Phi\left(\textbf{x},t\right)=\frac{1}{\sqrt{\lambda}}\Phi\left(x_1,x_2\right)\exp\left[i\left(p_3x_3-p_0t\right)\right],
\label{42}
\end{equation}
where $p_3=2\pi n_3/\lambda$, $n_3$ is a vertical quantum number, $\lambda$ is a cut-off of the integration over the $x_3$. Replacing Eq.(42) into Eq.(41), one finds
\[
\biggl[\eta\left(\partial_1^2+\partial_2^2\right)-e_0H\eta L_3-\frac{e_0^2H^2\eta}{4}\left(x_1^2+x_2^2\right)
\]
\vspace{-7mm}
\begin{equation} \label{43}
\end{equation}
\vspace{-7mm}
\[
-\eta p_3^2+p_0^2-m^2\biggr]\Phi\left(x_1,x_2\right)=0,
\]
where $\eta=1-Lp_0$. It is convenient to introduce cylindrical coordinates $x_1=r\cos\varphi$, $x_2=r\sin\varphi$, and then $L_3=-i\partial/\partial \varphi$. The solution to Eq.(43) in cylindrical coordinates exists in the form
\begin{equation}
\Phi\left(x_1,x_2\right)=\frac{\exp\left(il\varphi\right)}{\sqrt{2\pi}}\Psi(r),
\label{44}
\end{equation}
with $l$ being an orbital quantum number, $l=...,-2,-1,0,1,2,...$.  Introducing new variable $\rho=e_0Hr^2/2$, Eq.(43) in cylindrical coordinates, and with taking into account (44), becomes
\begin{equation}
\left(\rho\frac{d^2}{d\rho^2}+\frac{d}{d\rho}-\frac{l^2}{4\rho}-\frac{\rho}{4}+P\right)\Psi\left(\rho\right)=0,
\label{45}
\end{equation}
where $P=\left(p_0^2-\eta p_3^2-m^2-e_0Hl\eta\right)/\left(2e_0H\eta\right)$. The finite solution (at $\rho=0$ and $\rho\rightarrow \infty$) to Eq.(45) is given by \cite{Kruglov3}
\begin{equation}
\Psi(\rho)=\frac{N_0}{\sqrt{n!s!}}\exp\left(-\frac{\rho}{2}\right)\rho^{l/2}Q_s^l(\rho),
\label{46}
\end{equation}
where $N_0$ is the normalization constant, $s=0,1,2,...$ is the radial quantum number, $Q_s^l(\rho)$ is the Laguerre polynomial \cite{Bateman}
\begin{equation}
Q_s^l(\rho)=e^{\rho}\rho^{-l}\frac{d^s\left(\rho^{s+l}e^{-\rho}\right)}{d\rho^s}.
\label{47}
\end{equation}
The energy $p_0$ is quantized and is given by
\begin{equation}
p_0^2= \eta p_3^2+m^2+e_0H\eta\left(2n+1\right),
\label{48}
\end{equation}
where $n=l+s=0,1,2,...$ is a principal quantum number. The orbital quantum number runs the values $-\infty <l<n$. For negative orbital quantum number $l$, one can use the relation
\begin{equation}
\left(-1\right)^l\rho^{-l}Q_{s+l}^{-l}(\rho)= Q_s^l(\rho).
\label{49}
\end{equation}
Eq.(48) at $\eta=1$ ($L=0$) is converted into the known expression corresponding to the Klein$-$Gordon equation \cite{Sokolov}. Eq.(48) is consistent with Eq.(1) because as for the Klein$-$Gordon equation for scalar particles in external magnetic fields one has to make the replacement $p_1^2+p_2^2\rightarrow e_0H\left(2n+1\right)$ in the dispersion equation \footnote{Authors of \cite{Ellis2} obtained different from (48) expression because of their non-standard coupling with the electromagnetic fields.}. From (42),(44),(46), one obtains
\begin{equation}
\Phi\left(\textbf{x},t\right)=\frac{N_0}{\sqrt{n!s!}}\frac{e^{il\varphi}}{\sqrt{2\pi}}
\frac{\exp\left[i\left(p_3x_3-p_0t\right)\right]}{\sqrt{\lambda}}
e^{-\rho/2}\rho^{l/2}Q_s^l(\rho).
\label{50}
\end{equation}
The coefficient $N_0$ can be obtained from the normalization \cite{Sokolov}:
\begin{equation}
\frac{p_0}{m}\int_0^\infty rdr\int_{-\lambda/2}^{\lambda/2}dx_3\int_0^{2\pi}\Phi^\ast(x)\Phi(x)d\varphi =1.
\label{51}
\end{equation}
Calculating integrals in Eq.(51) with the help of the relation
\[
\int_0^\infty e^{-\rho}\rho^l\left[Q_s^l(\rho)\right]^2d\rho =s!\Gamma\left(l+s+1\right),
\]
where $\Gamma(x)$ is the Gamma function, and using the wave function (50), we find the normalization constant:
\begin{equation}
N_0 =\sqrt{\frac{e_0Hm}{p_0}}.
\label{52}
\end{equation}
Eqs.(48),(50),(52) allow us to investigate the synchrotron radiation of spinless particles with
modified dispersion relation (1). Charged particles moving in an external magnetic field (in helical orbits) emit the synchrotron radiation with frequency depending on the radius of the orbit \cite{Sokolov}. The orbit radius can be estimated by the classical relation \cite{Sokolov}:
\begin{equation}
R =\frac{\beta p_0}{e_0H},
\label{53}
\end{equation}
where $\beta=v$ is a particle velocity (in our notations $c=1$). Let us consider the case $p_3=0$ when particles rotate in cycles with ultra-relativistic energies ($p_0\gg m$, $v\approx 1$). Then Eq.(48)
becomes
\begin{equation}
p_0^2\approx e_0H\left(2n+1\right)\left(1-Lp_0\right).
\label{54}
\end{equation}
From quadratic equation (54), we obtain the approximate solution for positive energy implying that $Lp_0\ll 1$, $L\sqrt{e_0H\left(2n+1\right)}\ll 1$:
\begin{equation}
p_0\approx \sqrt{e_0H\left(2n+1\right)}-Le_0H\left(n+\frac{1}{2}\right).
\label{55}
\end{equation}
With the help of Eq.(55), we obtain from Eq.(53) the radius of the orbit
\begin{equation}
R \approx R_0-Ln,
\label{56}
\end{equation}
where
\begin{equation}
R_0 \approx\sqrt{\frac{2n}{e_0H}}
\label{57}
\end{equation}
is the radius of the orbit within the Klein$-$Gordon equation, and we took into consideration that for
ultra-relativistic energies $n\gg 1$, $n+1/2\approx n$. Eq.(56) indicates that the deformation parameter $L$ reduces the radius of the orbit ($L>0$). Thus, for high energies the second term in Eq.(56) should be taken into account. As a result, the angular orbital frequency $\omega=v/R$ is greater than in Lorentz-invariant theory, $\omega>\omega_0$ ($\omega_0=v/R_0$). To clear up the physical meaning of the radial quantum number $s$, we calculate the quantum average square radius
\begin{equation}
\overline{r^2}_{quant}=\frac{p_0}{m}\int \Psi^\ast(x)r^2\Psi(x)d^3x.
\label{58}
\end{equation}
Evaluating integral in Eq.(58) using the equality
\[
\int_0^\infty e^{-\rho}\rho^{l+1}\left[Q_s^l(\rho)\right]^2d\rho =n!s!\left(n+s+1\right),
\]
and $n=l+s$, one obtains \cite{Sokolov}
\begin{equation}
\overline{r^2}_{quant}=\frac{2}{e_0H}\left(n+s+1\right).
\label{59}
\end{equation}
Comparing the macroscopic classical average square radius \cite{Sokolov} $\overline{r^2_{cl}}=R^2+a^2$, where $a$ is the distance between the center of the trajectory and the origin, with the quantum average square radius (59), we find \cite{Sokolov}
\begin{equation}
a\approx\sqrt{\frac{2s}{e_0H}}.
\label{60}
\end{equation}
From Eqs.(57),(60), one obtains $l\approx e_0H(R_0^2-a^2)/2$ so that at $R_0>a$ the orbital quantum number is positive, $l>0$, and at $R_0<a$, we have $l<0$. One can construct the coherent states of a spinless particle at non-relativistic energy following the way of the work \cite{Kruglov3}.

To calculate the concrete numerical deformation effect, we use the possible energy of electrons $p_0=1$ TeV, and the magnetic field $H=260~\mu$G in Crab Nebula. The Heaviside$-$Lorentz system with the fine structure constant $\alpha=e^2/(4\pi)$ is used. In this system, the SI units are related to the energy units as follows: $1$ T$=195.5$ eV$^2$, $1$ m$=5.1\times10^6$ eV$^{-1}$. Then, one obtains the principal quantum number $n\approx p_0^2/(2e_0H) \approx 3\times 10^{29}$, and the change of the synchrotron radius (at $L=L_P$) $\Delta R=L_P n\approx
5~\mu$m. This is the very small amount compared to the classical radius (57). But if the deformation parameter $L$ is much greater than the Planck length $L_P$, then the deformation effect should be taken into consideration.

\section{Conclusion}

We have postulated the wave equation for spinless particles with the modified dispersion relation. Such dispersion relation is realized in DSR. The wave equation is formulated in the form of first-order $6\times 6$-matrix wave equation. The algebra of the $6\times6$-matrices of the equation has been obtained which is more complicated compared to the Duffin$-$Kemmer$-$Petiau algebra. We find the density matrix which can be used for different quantum field theory calculations. The Schr\"{o}dinger form of the equation in the $2\times 2$ matrix-differential form is obtained and quantum-mechanical Hamiltonian is found. The Hamiltonian obtained can be explored in quantum-mechanical evaluations. We find exact solutions to the wave equation for particles in constant and uniform external magnetic fields, and the synchrotron radius correction (due to quantum gravity) is estimated. Such solutions may be applied for the analysis of the synchrotron radiation from the Crab Nebula in the approximation when spin effects of electrons can be ignored. Thus, the bound on the deformation parameter $L$, within our approach, can be investigated. We leave the question about the invariance of Eq.(2) under DSR transformations and other problems for further learning.

\section{Appendix: Useful products of matrices}

With the help of Eq.(5), we obtain the products of $\beta$-matrices (7):
\begin{equation}
\beta_\mu\beta_\nu=\delta_{\mu\nu}\varepsilon^{0,0}+\varepsilon^{\mu,\nu}+\delta_{\nu 4}\varepsilon^{\mu,\widetilde{0}}-mL\left(\delta_{\mu m}\delta_{m\nu }\varepsilon^{\widetilde{0},0}+\delta_{\mu 4}\delta_{\nu m}\varepsilon^{0, m}\right),
\label{61}
\end{equation}
\[
\beta_\mu\beta_\nu\beta_\lambda=\delta_{\mu\nu}\left(\varepsilon^{0,\lambda}+\delta_{\lambda 4}\varepsilon^{0,\widetilde{0}}\right)
\]
\vspace{-7mm}
\begin{equation} \label{62}
\end{equation}
\vspace{-7mm}
\[
-mL\left[\delta_{\lambda m}\left(\delta_{m\nu}\delta_{\mu 4}\varepsilon^{0,0}
+\delta_{\nu 4}\varepsilon^{\mu,m}\right)+\delta_{\mu m}\delta_{m\nu}\left(\delta_{\lambda 4}\varepsilon^{\widetilde{0},\widetilde{0}}+\varepsilon^{\widetilde{0},\lambda}\right)\right],
\]
\[
\beta_\mu\beta_\nu\beta_\lambda\beta_\sigma=\delta_{\mu\nu}\delta_{\lambda\sigma}\varepsilon^{0,0}+
\delta_{\nu\lambda}\varepsilon^{\mu,\sigma}+ \delta_{\nu\lambda}\delta_{\sigma 4}\varepsilon^{\mu,\widetilde{0}}
\]
\begin{equation}
-mL\biggl[\delta_{\sigma m}\left(\delta_{m\lambda}\delta_{\nu 4}\varepsilon^{\mu,0}
+\delta_{\mu\nu}\delta_{\lambda 4}\varepsilon^{0,m}\right)
+\delta_{\mu 4}\delta_{\nu m}\delta_{m\lambda}\left(\varepsilon^{0,\sigma}+
\delta_{\sigma 4}\varepsilon^{0,\widetilde{0}}\right)
\label{63}
\end{equation}
\[
+\delta_{\mu n}\delta_{n\nu}\left(\delta_{\lambda\sigma}\varepsilon^{\widetilde{0},0}-
mL\delta_{\lambda 4}\delta_{\sigma m}\varepsilon^{\widetilde{0},m}\right)\biggr].
\]
From Eq.(7), we obtain the matrix $\hat{p}=p_\mu\beta_\mu$:
\begin{equation}
\hat{p}=p_\mu\left(\varepsilon^{\mu,0}+\varepsilon^{0,\mu}\right)+p_4\varepsilon^{0,\widetilde{0}}
-mLp_n\varepsilon^{\widetilde{0},n}.
\label{64}
\end{equation}
Taking into account Eqs.(61),(63), one finds
\begin{equation}
\hat{p}^2=p^2\varepsilon^{0,0}+p_\mu p_\nu\varepsilon^{\mu,\nu}+p_4p_\mu\varepsilon^{\mu,\widetilde{0}}
-mL\left(\textbf{p}^2\varepsilon^{\widetilde{0},0}+p_4p_n\varepsilon^{0,n}\right),
\label{65}
\end{equation}
\begin{equation}
\hat{p}^4=p^2\hat{p}^2
-mLp_4\textbf{p}^2\hat{p},
\label{66}
\end{equation}
where $p^2=p_\mu^2$. Useful relations for the matrix
\begin{equation}
\beta_4=\varepsilon^{4,0}+\varepsilon^{0,4}+\varepsilon^{0,\widetilde{0}}
\label{67}
\end{equation}
are as follows:
\begin{equation}
\beta_4^3=\beta_4,~~~~\beta_4^2=\varepsilon^{0,0}+\varepsilon^{4,4}+\varepsilon^{4,\widetilde{0}},
\label{68}
\end{equation}
\begin{equation}
\beta_4\beta_m=\varepsilon^{4,m}-mL\varepsilon^{0,m},
\label{69}
\end{equation}
\begin{equation}
\beta_4\beta_m\beta_4^2=0,~~~~~~ \beta_4\beta_m\left(I_6-\beta_4^2\right)\beta_n\left(I_6-\beta_4^2\right)=0.
\label{70}
\end{equation}

\end{document}